\documentstyle[epsf,preprint]{ptptex}
\preprintnumber[3cm]{
NUP-A-2002-7\\
}
\title{5-Dimensional Spacetime with $q$-Deformed Extra Dimension}

\author{
Shigefumi {\sc Naka} and
Haruki {\sc Toyoda}
}

\inst{
Department of Physics, College of Science and Technology\\
Nihon University, Tokyo 101-8308,Japan \\
(Received September 11, 2002)
}

\abst{
An attempt to obtain the non-trivial mass structure of particles in a Randall-Sundrum type of 5-dimensional spacetime with a q-deformed extra dimension is discussed. In this spacetime, the five-dimensional space has no boundary, but there arises an elastic potential preventing free motion in the fifth direction. The q-deformation is, then, introduced in such a manner that a non-commutativity arises in the spacetime coordinates between the 4-dimensional components and the fifth component. As a result of this q-deformation, the propagators of particles embedded in this spacetime naturally acquire an ultraviolet-cutoff effect without spoiling the Lorentz covariance in the 4-dimensional spacetime.
}

\begin{document}

\maketitle

\section{Introduction}

Recently, the role of extra dimensions has been actively investigated by many people from new points of view. One approach to understanding this role proposed by Randall-Sundrum is intended for deriving a large mass hierarchy from a 5-dimensional spacetime model. \cite{Randall} In such a model, the scale of the extra dimensions is specified by periodic boundary conditions or by some compactness conditions. Although the physical reality of the thickness of the extra dimension and its stability open questions, this type of model possesses some advantages over the standard compactifications, such as the Kaluza-Klein theory. 

 As shown by Connes, a non-commutative geometry associated with some types of discrete-extra dimensions leads to a geometric interpretation of the Standard Model, including Higgs fields.\cite{Connes} Furthermore, it is also found that D-brane models with a constant B-field \cite{Callan} lead to the non-commutativity of 4-dimensional spacetime coordinates. In the latter case, field theories on such spacetimes have an interesting property relating UV divergence to IR divergence.\cite{Seiberg,Kaminsky}

In this paper, we study a different type of non-commutativity based on a $q$-deformed quantization\cite{Macfarlane} of a 5-dimensional spacetime, which is similar to the Randall-Sundrum model. In our model, the fifth dimension is obtained as a hard elastic continuum without boundaries. Particles embedded in this spacetime can move freely in the directions of 4-dimensional spacetime, but they experience very strong forces opposing their motion in the fifth direction. After setting up the 5-dimensional spacetime, we further attempt to deform the spacetime to cause mixing of phase spaces between the 4-dimensional spacetime and the extra space. In the resultant spacetime, the 4-dimensional coordinates remain commutative; it is interesting, however, that fields embedded in this spacetime exhibit desirable UV behavior.

In the next section, first, we derive the five-dimensional spacetime with an elastic fifth dimension from a Randall-Sundrum type of metric. Then, discussion is given of the non-commutative structure from the viewpoint of the {\it q}-deformation.

In $\S$ 3, the properties of a scalar field embedded in such a 5-dimensional spacetime are studied. There, the mass eigenvalue obtained from the oscillation mode in the fifth dimension is discussed, in addition to the UV behavior of a loop graph. Section 4 is devoted to a summary of characteristics of the present model with discussion of remaining problems. We also give a short review of $q$-oscillator variables in Appendix A. In Appendix B, the relation between the Laplacian and the scalar curvature in a spacetime that is manifestly conformal to {\it D}-dimensional Minkowski spacetime is studied. Appendix C presents a supplementary explanation for the loop calculation given in $\S$ 3.

\section{A model of 5-dimensional spacetime}

The basic metric structure of the 5-dimensional spacetime in our model is essentially the same as the Randall-Sundrum type of spacetime except that the extra dimension has no boundary. The line element in our model is given by
\begin{equation}
 ds^2=g_{ij}dx^idx^j=e^{-2\sigma(\eta)}\eta_{\mu\nu}dx^\mu dx^\nu-d\eta^2,~(\mu,\nu=0,1,2,3;~dx^5=d\eta ) \label{metric}
\end{equation}
where $(\eta_{\mu\nu})={\rm diag}(+---)$; more precisely, we consider the simple case that the metric of the 4-dimensional spacetime is flat Minkowski spacetime. The spacetime is static, and they are, non-vanishing components of the Christoffel symbols become
\begin{equation}
 \Gamma^\mu_{5\nu} = -\sigma^\prime\delta^\mu_\nu, ~~ \Gamma^5_{\mu\nu} = -\sigma^\prime\eta_{\mu\nu}e^{-2\sigma} . \label{Christoffel}
\end{equation}
Using these expressions for $\Gamma^i_{jk}$, we can calculate the Ricci tensor and the scalar curvature, respectively, in the following forms:
\begin{eqnarray}
 R_{\mu\nu} &=& \eta_{\mu\nu}e^{-2\sigma}(-\sigma^{\prime\prime} + 4\sigma^{\prime 2}) , \\
 R_{55} &=& 4(\sigma^{\prime\prime}-\sigma^{\prime 2}) , \\
 R &=& -8\sigma^{\prime\prime}+20\sigma^{\prime 2} .
\end{eqnarray}
Here, the prime denotes differentiation with respect to $\eta$. To determine the function $\sigma(\eta)$, we consider the case
\begin{equation}
 S \sim \int d^5 x \sqrt{g}(R+2\Lambda)
\end{equation}
as the action for the 5-dimensinal spacetime, where $\Lambda$ is a cosmological constant in the 5-dimensional spacetime. We require $\delta S=0$ under variations, so that
\begin{equation}
 \delta g_{\mu\nu}=-2\delta\sigma g_{\mu\nu},~\delta g_{55}=0 .
\end{equation}
Then, we have the vacuum equations $R_{\mu\nu}-\frac{1}{2}g_{\mu\nu}(R+2\Lambda)=0$, or
\begin{equation}
 \sigma^{\prime\prime}-2\sigma^{\prime 2}-\frac{\Lambda}{3}=0. \label{Einstein eq}
\end{equation}
It should be that the function $\sigma(\eta)$ can be determined only by Eq.\,(\ref{Einstein eq}), and the condition $\delta g_{55}=0$ ensures the form invariance of the metric.
\footnote{
The effective Lagrangian $L$ of $\sigma$ should be defined by $S \sim \int d\eta L(\sigma,\sigma^\prime)$, which gives rise to $ L=\frac{1}{2}e^{-4\sigma}\left(\sigma^{\prime 2}-\frac{\Lambda}{6}\right) $. Then, the Lagrange equation for $\sigma$ coincides with Eq.\,(\ref{Einstein eq}).
}

Equation (\ref{Einstein eq}) possesses several solutions, and for $\Lambda <0$, we obtain
\begin{equation}
 \sigma(\eta)=-\frac{1}{2}\log\sinh\left(\sqrt{-\frac{2\Lambda}{3}}\eta\right)+{\rm const} , \label{sigma 1}
\end{equation}
which is suitable for later use. We fix the additional constant in (\ref{sigma 1}) by requiring $e^{-2\sigma(\eta)} \rightarrow 2\kappa\eta~(\Lambda \rightarrow 0)$, where $\kappa$ is a constant with dimension $\eta^{-1}$. Therefore, a possible form of $\sigma$ is
\begin{equation}
 \sigma(\eta)=-\frac{1}{2}\log\left\{ 2\kappa\sqrt{-\frac{3}{2\Lambda}}\sinh\left(\sqrt{-\frac{2\Lambda}{3}}\eta\right) \right\}.
\end{equation}

The next step is to introduce the variable $y$ defined by $d\eta=e^{-\sigma}dy$.  Then, the metric in $(x^\mu, y)$ coordinates becomes manifestly conformal to that in 5-dimensional Minkowski spacetime, and we obtain
\begin{equation}
 g_{ij}=e^{-2\sigma}\eta_{ij},~\left( ~(\eta_{ij})=diag(+----)~ \right) \label{conformal-flat}
\end{equation}
where $\sigma(\eta(y))$ is a function of $y$ through $\eta$. The function $\eta(y)$ can't be obtained in a simple functional form except in  the case $\Lambda \rightarrow 0$. In this exceptional case, $\eta$ can be obtained as $\eta=\frac{\kappa}{2}y^2$, and then, we have
\begin{equation}
 e^{-2\sigma(\eta(y))}=2\kappa\eta(y)=\kappa^2 y^2 . \label{sigma 2}
\end{equation}
In order to understand the role of the metric (\ref{sigma 2}) in particle physics, let us consider a classical free particle embedded in this spacetime. The action for a particle of mass $m$ should be
\begin{equation}
 S=-m\int d\tau\sqrt{g_{ij}\dot{x}^i\dot{x}^j}, \label{particle}
\end{equation}
where the dot denotes differentiation with respect to the time ordering parameter $\tau$. The momentum conjugate to $x^i$ has the form
\begin{equation}
 p_i=\frac{\delta S}{\delta \dot{x}^i}=-m\sqrt{e^{-2\sigma}}\frac{\eta_{ij}\dot{x}^j}{\sqrt{\dot{x}^2}},~(\dot{x}^2=\eta_{ij}\dot{x}^i\dot{x}^j)
\end{equation}
from which we can obtain the mass-shell equation\footnote{
If necessary, one can use the action $S'=-\frac{1}{2}\int d\tau\left(\frac{1}{e}g_{ij}\dot{x}^i\dot{x}^j+m^2 e\right)$ instead of (\ref{particle}). Here, $e(\tau)$ is the einbein, by which $d\tau e$ is made invariant under the reparametrization of $\tau$. The action $S'$ allows the massless limit $m\rightarrow 0$, and it leads to the same constraint as (\ref{mass-shell-1}) by varying $S'$ with respect to $e$.}
\begin{equation}
 \eta^{ij}p_i p_j - m^2e^{-2\sigma}=p^2- \{ p_y^2+(m\kappa)^2 y^2 \} =0 , \label{mass-shell-1}
\end{equation}
with $p^2=p^\mu p_\mu$. The mass term of the particle in 4-dimensional spacetime is identical to the Hamiltonian of a harmonic oscillator in the five-dimensional space. In $q$-number theory, the mass-shell equation can be regarded as the wave equation for a particle embedded in this spacetime,
\begin{equation}
 \left( p^2-m\kappa \{ a,a^\dagger \} \right) \Psi =0, \label{mass-shell-2}
\end{equation}
where $a$ and $a^\dagger$ are the oscillator variables defined by
\begin{equation}
 a=\frac{1}{\sqrt{2m\kappa}}(m\kappa y+ip_y),~ a^\dagger =\frac{1}{\sqrt{2m\kappa}}(m\kappa y-ip_y). \label{oscillator}
\end{equation}
In other words, the particles embedded in this spacetime acquire a linearly increasing mass spectrum, with the ground-state mass $\sqrt{m\kappa}$, except in the case $m=0$. In the exceptional case, all spectrum degenerate into a massless state. As is discussed in the next section, the form of Eq.\,(\ref{mass-shell-2}) is not completely correct as a field equation in a curved $(x^\mu)$ spacetime, because the problem of ordering between $g_{yy}$ and $p_y$ has been d ignored.

Now, as the third step in the model construction, we require a $q$-deformation between $x^\mu$ and $y$ in the mass-square term in Eq.\,(\ref{mass-shell-2}). Usually, a $q$-deformation is defined by a modified commutation relation including a parameter $q$. Then, the spectrum of the $q$-deformed harmonic oscillator is modified from that of equally spaced eigenvalues. The manner of the $q$-deformation, however, is not unique; in what follows, we shall consider the case defined by the replacement $(a,a^\dagger) \rightarrow (a_q,a_q^\dagger)$ with
\begin{equation}
 a_q = a \sqrt{\frac{[N]_q}{N}},~a_q^\dagger = \sqrt{\frac{[N]_q}{N}}a^\dagger,
\label{q-deformation-1}
\end{equation}
where
\begin{equation}
 N=a^\dagger a~~{\rm and}~~[N]_q=\frac{\sinh[\alpha (N + \gamma p^2)\log q]}{\sinh(\alpha\log q)} . \label{q-deformation-2}
\end{equation}
Equation. (\ref{q-deformation-1}) with (\ref{q-deformation-2}) is a possible solution of the $q$-deformed commutation relation (Appendix A)
\begin{equation}
 a_q a^\dagger_q-q^\alpha a^\dagger_q a_q =q^{-\alpha\left(N +\gamma p^2\right) }, \label{q-deformation-3}
\end{equation}
where $\alpha, \gamma$ and $p^2$ are deformation parameters in this case. By including $p^2$ in the deformation parameters, the substitution of $(a_q, a_q^\dagger)$ for $(a,a^\dagger)$ causes mixing between the 4-dimensional spacetime and the extra space. Explicitly, with $[x^\mu,p^\nu]=-i\eta^{\mu\nu}$, the commutator of $x^\mu$ and $y_q=\frac{1}{\sqrt{2m\kappa}}(a^\dagger_q + a_q)$ becomes non-vanishing,
\begin{equation}
 [x^\mu,y_q]=-i(\alpha\gamma\log q)p^\mu\frac{1}{\sqrt{2m\kappa}}\left(\coth[\alpha(N+\gamma p^2)\log q]a^\dagger_q + {\rm h.c.} \right),
\end{equation}
provided that $\alpha\gamma \neq 0$.

Therefore, the mass-shell equation after the $q$-deformation should be written
\begin{equation}
 \left( p^2-m\kappa \{ a_q,a_q^\dagger \} \right) \Psi =0 . \label{mass-shell-3}
\end{equation}
Although it is a logical jump to replace Eq.\,(\ref{mass-shell-2}) with Eq.\,(\ref{mass-shell-3}), the dynamics described by the wave equation after the $q$-deformation have interesting properties. In particular, the mass-square-like term in Eq.\,(\ref{mass-shell-3}) is a $p^2$-dependent operator:
\begin{equation}
 {\cal M}^2(p^2) = m\kappa\{a_q,a_q^\dagger\} = m\kappa \frac{\sinh \left[\alpha(\frac{1}{2}\{ a,a^\dagger\}+\gamma p^2)\log q \right]}{\sinh(\frac{1}{2}\alpha\log q)}.   \label{q-mass-square}
\end{equation}
The mass eigenvalues are, thus, determined as solutions of the self-consistent equation $M^2={\cal M}^2(M^2)$ for each eigenvalue of $N$. As shown in Fig.\,1,  the solutions appear at the intersections of $x$ and ${\cal M}^2(x)$. Thus, the mass-square spectrum undergoes a modification from a linear function of $N$ to a non-linear one characterized by the parameters $\alpha$ and $\gamma$.

\begin{figure}[t] 
 \parbox[t]{\halftext}{
  \epsfxsize=5cm \epsfbox{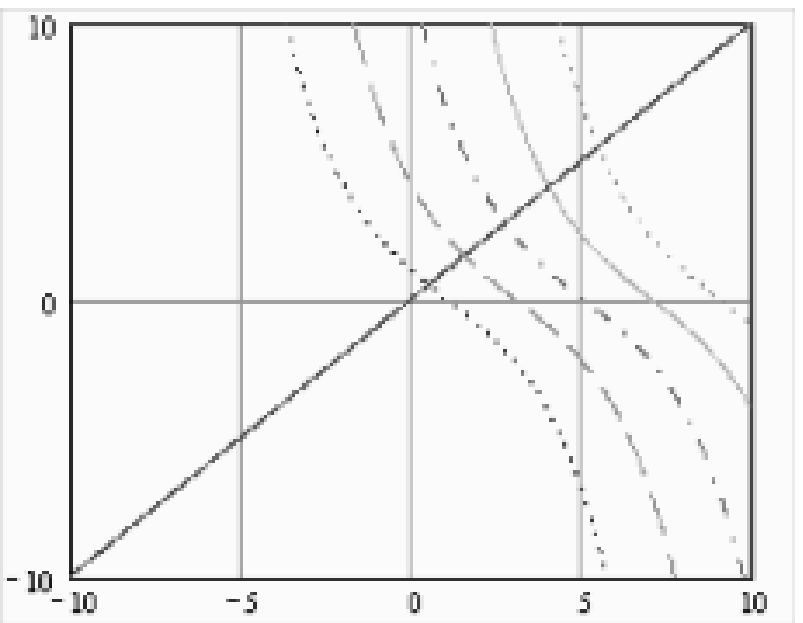} 
    \caption{Mass eigenvalues appear at the intersections of the straight line and curved lines corresponding to $N=0,1,2,\cdots$ with $\gamma < 0$.}
                      }
\hspace{8mm}
 \parbox[t]{\halftext}{
  \epsfxsize=5cm \epsfbox{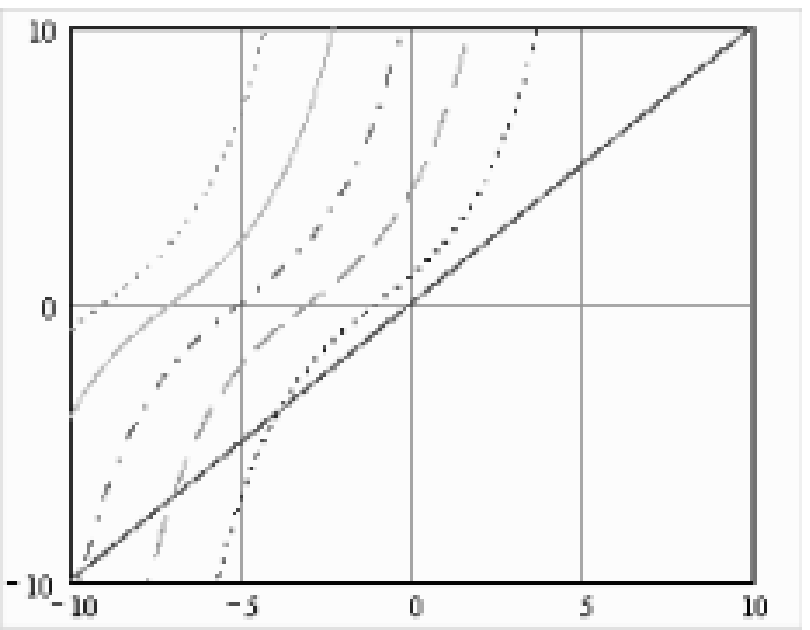} 
    \caption{For $\gamma > 0$, only space-like solutions, $M^2<0$, are allowed. }
                      }
\end{figure}

We also note that in the modified field equation (\ref{mass-shell-3}), the term ${\cal M}^2(p^2)$ is a function of $(p^2+\frac{1}{2}\gamma^{-1}\{a,a^\dagger\})$, which can be seen to have the same structure as Eq.\,(\ref{mass-shell-2}) by reading $-\gamma$ as ($2m\kappa)^{-1}$. In this sense, it can be said that the modified field equation exhibits fractal structure with respect to the scale of $p^2$. Another interesting result of the deformation (\ref{q-deformation-1}) with (\ref{q-deformation-2}) is that the propagator, the inverse of the modified field equation, rapidly decreases as $p^2$ increases, for this reason, good convergent properties are guaranteed for some loop diagrams.

\section{Scalar field embedded in the five-dimensional spacetime}

In this section, we study a scalar field embedded in the five-dimensional spacetime $(x^i)=(x^\mu,y)$ characterised by the metric $g_{ij}=e^{-2\sigma}\eta_{ij}~(d\eta =e^{-\sigma}dy)$ with vanishing 5-dimensional cosmological constant $\Lambda=0$. In this case, the action of the scalar field should be
\begin{equation}
 S=\int d^5x\sqrt{g}\left[ \frac{1}{2}\left\{ g^{ij}\partial_i\varphi\partial_j\varphi - (m^2-\xi R)\varphi^2 \right\} - V(\varphi) \right] , \label{action1}
\end{equation}
where $\xi$ is a parameter adjusting the conformal invariance of free field theory. Varying the action with respect to $\varphi$, we have (Appendix B)
\begin{equation}
  \left[ e^{2\sigma}\left(\partial_i-\frac{3}{2}\partial_i\sigma \right)^2 + m^2 + \left( \frac{3}{16}-\xi \right) R \right]\varphi + V^\prime(\varphi)=0 .
\end{equation}
Carrying out, here, the scale transformation of the scalar field according to
\begin{equation}
 \varphi=e^{\frac{3}{2}\sigma}\tilde{\varphi}, \label{scale}
\end{equation}
the field equation for $\tilde{\varphi}$ becomes
\begin{equation}
 \left[ \partial_\mu\partial^\mu - \partial_y^2 + e^{-2\sigma}\left\{ m^2 + \left( \frac{3}{16}-\xi \right) R \right\} \right]\tilde{\varphi} + e^{-\frac{7}{2}\sigma}V^\prime(e^{\frac{3}{2}\sigma}\tilde{\varphi})=0  . \label{eq. for varphi}
\end{equation}
Because we are considering the case $\Lambda=0$, the scalar curvature can be calculated as ${\displaystyle R=\eta^{-2}=4(\kappa y^2)^{-2} }$, with $\sigma=-\log|\kappa y|$ in Eq.\,(\ref{sigma 2}). The scalar curvature, then, adds a potential term in $y$ space other than $e^{-2\sigma}m^2$, which can be removed by choosing $\xi=\frac{3}{16}$. Then, the free field equation with $V=0$ can be reduced to Eq.\,(\ref{mass-shell-1}), that is, the mass-shell equation in flat 4-dimensional spacetime having a harmonic oscillator in $y$ space as its mass-square term.

Refurning to the case $V \neq 0$, let us carry out the scale transformation (\ref{scale}) in the action (\ref{action1}). We obtain
\begin{equation}
 S=\int d^5x \left[ \frac{1}{2}\left\{ \eta^{ij}\partial_i\tilde{\varphi} \partial_j\tilde{\varphi}-e^{-2\sigma}\left\{m^2+\left(\frac{3}{16}-\xi\right) R \right\}\tilde{\varphi}^2 \right\} - e^{-5\sigma}V \left( e^{\frac{3}{2}\sigma}\tilde{\varphi} \right) \right] . \label{action2}
\end{equation}
Equation.\,(\ref{eq. for varphi}) can be derived directly by varying the action (\ref{action2}) with respect to $\tilde{\varphi}$. Hereafter, we consider the non-singular case $\xi=\frac{3}{16}$ with $\sigma=-\log|\kappa y|$. Then, the $R$ term vanishes, and, in terms of oscillator variables, the action for $\tilde{\varphi}$ becomes
\begin{equation}
 S=\int d^5x \left[ -\frac{1}{2}\tilde{\varphi}\left\{ \partial_\mu \partial^\mu +(m\kappa) \{ a,a^\dagger \} \right\} \tilde{\varphi} - |\kappa y|^5V \left( |\kappa y|^{-\frac{3}{2}}\tilde{\varphi} \right) \right], \label{action3}
\end{equation}
after the integration by parts. It should be noted that the Heisenberg equation for $\tilde{\varphi}$ is regular at $y=0$, provided that $V(\varphi)$ is a polynomial up to quartic degree in $\varphi$, even if the potential contains a singularity at $y=0$. The last step in building our model is to deform the oscillator variables using Eqs.\,(\ref{q-deformation-1}) and (\ref{q-deformation-2}) so that $x^\mu$ and $(a_q,a_q^\dagger)$ become non-commuting variables. More precisely, we substitute the action (\ref{action3}) for the $q$-deformed action defined by
\begin{equation}
 S_q=\int d^5x \left[ -\frac{1}{2}\tilde{\varphi}\left\{ \partial_\mu \partial^\mu +(m\kappa) \{ a_q,a_q^\dagger \} \right\} \tilde{\varphi} - |\kappa y|^5V \left( |\kappa y|^{-\frac{3}{2}}\tilde{\varphi} \right) \right]. \label{action4}
\end{equation}
Here, we have regarded the $q$-deformation as the mapping in the phase space $(y,p_y)$ defined by $p_y^2 \rightarrow m\kappa \{ a_q,a_q^\dagger \}-(m\kappa)^2y^2,~ y \rightarrow y$, so that the form of potential remains invariant under this deformation.

To study field theoretical properties of the system described by the action (\ref{action4}), let us consider the following simple case as an example:
\begin{equation}
 \int d^5x|\kappa y|^5V\left(|\kappa y|^{-\frac{3}{2}}\tilde{\varphi}\right)= \frac{\lambda}{4!}\int d^5x\rho(y)\tilde{\varphi}^4 ~.
\end{equation}
Writing the Klein-Gordon operator in $\{x^i\}$ space as $K=\partial_\mu\partial^\mu+(m\kappa)\{a_q,a^\dagger_q\}$, the Feynman propagator $G_{12}=\langle 0|T\tilde{\varphi}_1\tilde{\varphi}_2|0 \rangle$ gives rise to (Appendix C)
\begin{equation}
(iG)^{-1}_{12}=\left( K+\frac{\lambda}{2}\rho\bar{G}^{(0)}+O(\lambda^2) \right)_{12} , \label{Green function}
\end{equation}
where $(\bar{G}^{(0)})_{12}=-i(K^{-1})_{11}\delta_{12}$, and the indices $1$ and $2$ designate the two world coordinates $(x^\mu_i,y_i)~(i=1,2)$. Thus, defining the internal quantum number by $N|n \rangle=n|n\rangle$, the first-order-loop correction $\delta m^2 \equiv \frac{\lambda}{2}\rho\bar{G}^{(0)}$ to the mass square operator $(m\kappa)\{a_q,a_q^\dagger\}$ can be evaluated as $(\delta m^2)_{ij}=-i\frac{\lambda}{2}\rho_i(K^{-1})_{ii}\delta_{ij}$. Here,
\begin{eqnarray}
 (\bar{K}^{-1})_{ii} &=& \sum_n |\langle y_i|n\rangle |^2 \langle x^\mu_i,n|K^{-1}|x^\mu_i,n \rangle \nonumber \\
 &=&
 -\sum_n |\langle y_i|n\rangle|^2 \int\frac{d^4p}{(2\pi)^4}\frac{1}{p^2-m\kappa\frac{\sinh[\alpha(n+\frac{1}{2}+\gamma p^2)\log q]}{\sinh(\frac{1}{2}\alpha\log q)} } \nonumber \\
 &\sim& \pm \sum_n |\langle y_i|n\rangle|^2 \frac{2\sinh(\frac{1}{2}\alpha\log q)}{m\kappa(2\pi)^4}e^{\mp (n+\frac{1}{2})(\alpha\log q)} \left(\frac{\pi}{\alpha\gamma\log q}\right)^2 , \label{K inverse}
\end{eqnarray}
where $\pm$ is the signature of $\alpha\gamma\log q$. Although the relation "$\sim$" results from a rough estimation in the region $|p^2| \gg |\alpha\gamma\log q|$, it should be remarked that the integral is convergent, provided that  $\alpha\gamma\log q\neq 0$. This is due to the fact that the denominator of the integrand rapidly increases as $|p^2|$ increases. In consideration of the pole property of $K$ discussed in the previous section, therefore, it can be said that the above $\tilde{\varphi}^4$-type of interacting field has good convergent properties for $\gamma<0$ and $\alpha\log q>0$, that is, the case of $-$ signature. Then, the summation over $n$ in Eq.\,(\ref{K inverse}) converges, because $\langle y_i|n\rangle$ is a plynomial of degree $n$.
\footnote{
If we read $i$ as the sate $|x^\mu_i,n_i\rangle$ instead of $|x^\mu_i,y_i\rangle$, then $\delta m^2_{ij} \propto \sum_n\langle h_i h_n \rho h_n h_j\rangle e^{-(n+\frac{1}{2})(\alpha\log q)}$, where $h_i(y)=h_{n_i}(y)=\langle y|n_i\rangle$, and $\langle \cdots \rangle$ represents integration over $y$. In this case, the summation over $n$ is still convergent for a wide class of $\rho(y)$ including $\rho(y)=(\kappa y)^{-1}$,i.e.,the case $V(\varphi)=\varphi^4$.
}

\section{Summary and discussion}

In this paper, we have mainly discussed two characteristic points of a 5-dimensional spacetime model with a $q$-deformed extra space. First, the spacetime defined by the metric (1) with vanishing 5-dimensional cosmological constant allows structure of the fifth dimension that adds a harmonic-oscillator type of mass-square term to particles embedded in this spacetime. The fifth dimension is static and boundary free; that is, the extra dimension is non-compact. However, we do not need to worry about the existence of the fifth dimensions, because by adjusting $\kappa$, it becomes very difficult to move in its direction due to the existence of a strong force directed foward the origin $y=0$.

Second, the oscillator variables arising in the fifth dimension are deformed in such a way that $\partial_\mu\partial^\mu$ is included in the set of $q$-deformation parameters. As a result of this deformation, $\{x^\mu\}$ and $y$ become non-commuting, though $\{x^\mu\}$ remains as a set of commuting coordinates. It is interesting to compare this type of non-commutativity to that in D-brane models with a constant B-field background. In those models, the constant B-field results in non-commuting 4-coordinates, with $[x^\mu,x^\nu]=i\theta^{\mu\nu}$, where $\theta^{\mu\nu}$ is a constant antisymmetric tensor. Then, the field theories in that spacetime,i.e. non-commutative field theories, have different properties from the commutative field theories in the sense that an effective cutoff of one-loop nonplanar graphs replaces the UV divergence with an IR divergence associated with external momenta, though the Lorentz covariance of the theory is broken by $\theta^{\mu\nu}$. The cutoff property and the breaking of the Lorentz covariance may be complementary in those field theories.

By contrast, the non-commutativity in the present model can preserve the Lorentz covariance, and, further, some one-loop graphs become convergent as should be the case. We also emphasize that as illustrated in Figs. 1 and 2, the free propagator $K^{-1}(p^2)$ has simple poles $p^2$ at mass-square eigenvalues $m_0^2,\,m_1^2,\cdots ,$ such that $K(p^2)\simeq -(p^2-m_n^2)\left[1-m\kappa\alpha\gamma\frac{\cosh\{\alpha(n+\frac{1}{2}+\gamma m_n^2)\log q\}}{\sinh(\frac{1}{2}\alpha\log q)} \right]$ for $p^2\simeq m_n^2$. In other words, all residues of these poles have the same sign, and so there are, the present model is free from the problem of multi-pole ghosts. This situation is fairly different from that of the usual non-local field theories that contain higher derivative terms in the form of meromorphic functions.\cite{Pais}

The mass eigenvalues after the $q$-deformation are obtained by solving Eq.\,(\ref{q-mass-square}), where the quantum numbers of those eigenvalues may be read as generation. In order to estimate the scale of the mass eigenvalues, let us consider the case in which the $q$-deformation parameter $\alpha\log q$ is a small quantity, on the order of or smaller than $0.1$. Further, if we assume, for example, that the mixing between $\{x^\mu\}$ and $y$ due to the parameter $\gamma$ becomes effective at the mass scale $\kappa$\,( that is, $|\gamma|\kappa^2 \sim 1$), then the ground state mass is numerically estimated as $m_0^2 \sim m\kappa$. This result is, however, dependent on the choice of these parameters; if we assume a larger mixing parameter $|\gamma|$, then the ratio $m\kappa/m_0^2$ becomes several power of $10$. Therefore, we can't give a precise solution to the hierarchy problem.

In spite of these characteristic features of the present model, there is no theoretical background to introduce the $q$-deformation (\ref{q-deformation-1}),(\ref{q-deformation-2}) or (\ref{q-deformation-3}). Geometrical approaches to $q$-deformed phase space are known only for limited cases,\cite{Wess} and at the present stage, we do not have a good explanation for Eq.\,(\ref{q-deformation-2}) from such a geometrical point of view. Apart from the present 5-dimensional model, however, the resultant Eq.\,(\ref{mass-shell-3}) has a noteworthy structure, which may be applicable to some types of non-local field theories, e.g. bi-local field theories\cite{Yukawa} possessing a harmonic-oscillator type of mass-square term. Indeed, it has been attempted to relate the non-local structure of a field to the geometric structure of its momentum space.\cite{Snyder} These are interesting subjects for a future study.

\section*{Acknowledgements}

The authors wish to express their thanks to the members of their laboratory for discussions and encouragement. They are also grateful to one of the referees for his helpful comments. Equations (\ref{Green function}) and (\ref{K inverse}) were corrected using his suggestions.

\appendix

\section{Representation of a q-Oscillator}

We here study the representation of the $q$-oscillator variables satisfying

\begin{equation}
 a_q a_q^\dagger -q^\alpha a_q^\dagger a_q =q^{-\alpha ( N + \beta) } \label{q-commutator}
\end{equation}
in terms of the ordinary oscillator variables defined by $[a,a^\dagger]=1$. Here, $\alpha,\beta$ and $\lambda$ are parameters, and $N=a^\dagger a$. To find the representation, let us set
\begin{equation}
 a_q = a \sqrt{\frac{[N]_q}{N}},~~a_q^\dagger = \sqrt{\frac{[N]_q}{N}}a^\dagger ,
\end{equation}
where $N=a^\dagger a$ and $[N]_q$ is a function of $N$ determined below. Then, recalling that $Na=a(N-1)$ and $Na^\dagger=a^\dagger(N+1)$, we can verify
\begin{equation}
 a_q a_q^\dagger = [N+1]_q,~~a_q^\dagger a_q = [N]_q . \label{A1}
\end{equation}
Then Eq.\,(\ref{q-commutator}) can be reduced to
\begin{equation}
 [N+1]_q - q^\alpha [N]_q = q^{-\alpha (N +\beta) } .  \label{recurrence}
\end{equation}
The recurrence equation (\ref{recurrence}) can be solved easily as
\begin{equation}
 [N]_q = q^{-\alpha\beta } \frac{q^{\alpha N} - q^{-\alpha N}}{q^\alpha - q^{-\alpha}} + q^{\alpha N}[0]_q ,
\end{equation}
where $[0]_q$ is an arbitrary first term. Here, we choose
\begin{equation}
 [0]_q= \frac{q^{\alpha\beta}-q^{-\alpha\beta}}{q^\alpha - q^{-\alpha}} .
\end{equation}
Then, $[N]_q$ has the simple form
\begin{equation}
 [N]_q = \frac{q^{\alpha(N+\beta)}-q^{-\alpha(N+\beta)}}{q^\alpha -q^{-\alpha}} \label{A2}
\end{equation}
After this $q$-deformation, the Hamiltonian of the oscillator $\frac{1}{2}\{ a,a^\dagger \}$ should be replaced with $\frac{1}{2} \{ a_q,a^\dagger_q \}$, for which we have the expression
\begin{eqnarray}
 \frac{1}{2} \{ a_q,a^\dagger_q \} &=& \frac{1}{2}\left( [N]_q + [N+1]_q \right) \nonumber \\
  &=& \frac{1}{2}q^\lambda \frac{\sinh \left[\alpha(N+\frac{1}{2}+\beta)\log q \right]}{\sinh(\frac{1}{2}\alpha\log q)} \label{A3}
\end{eqnarray}

The $q$-deformation for the oscillator variables $(a,a^\dagger) \rightarrow (a_q,a_q^\dagger)$ can also be formulated as a result of the $q$-deformation for the phase space variables $(y,p_y) \rightarrow (y_q,p_{qy})$. Here, the $q$-deformed phase variables $(y_q,p_{qy})$ should be required to satisfy
\begin{equation}
 \frac{\omega}{2} \{ a_q,a^\dagger_q \}=\frac{1}{2}(p_{qy}^2+\omega^2 y_q^2) ,
\end{equation}
where $\omega$ is a constant, corresponding to $m\kappa$ in the previous sections. One way to define $(y_q,p_{qy})$ is to set
\begin{equation}
 y_q= \sqrt{\frac{1}{2\omega}}(a_q^\dagger + a_q),~~p_{qy}=i\sqrt{\frac{\omega}{2}}(a_q^\dagger - a_q).
\end{equation}
Another way is to set $p_y^2 = m\kappa \{ a_q,a_q^\dagger \}-(m\kappa)^2y^2,~ y_q =y$, as in \S 3.

\section{Laplacian in a D-dimensional Conformally Flat Spacetime}

We here demonstrate show the relation between the Laplacian and the scalar curvature in a spacetime that is manifestly conformal to $D$-dimensional Minkowski spacetime; that is, that we consider the spacetime with the metric
\begin{equation}
 g_{ij}(x)=e^{-2\sigma(x)}\eta_{ij}.~(i,j=0,1,\cdots,D-1; (\eta_{ij})={\rm diag}(+--\cdots -))
\end{equation}
The Christoffel symbol can be easily calculated from $g_{ij}$ as follows:
\begin{equation}
 \Gamma^i_{jk}=-(\partial_j\sigma)\delta^i_k-(\partial_k\sigma)\delta^i_j+(\partial^i\sigma)\eta_{jk}.~(\partial^j=\eta^{jk}\partial_k)
\end{equation}
From this expression, we can obtain the scalar curvature with a short calculation in the form
\begin{equation}
 R=(D-1)e^{2\sigma}\left[ 2(\partial^2\sigma)-(D-2)(\partial_i\sigma)(\partial^i\sigma) \right] .
\end{equation}

Now, the Laplacian in this spacetime is given by
\begin{eqnarray}
 \Delta &=& \frac{1}{\sqrt{|g|}}\partial_i\left(\sqrt{|g|}g^{ij}\partial_j\right) = e^{D\sigma}\partial_i\left( e^{-(D-2)\sigma}\partial^i \right)  \nonumber \\
 &=& e^{2\sigma}\left(\partial^2-(D-2)(\partial_i\sigma)\partial^i \right) . \label{Laplacian}
\end{eqnarray}
To remove the $(\partial_i\sigma)\partial^i$ term in the right-most term, we recall that
\begin{equation}
 \left(\partial_i-\frac{D-2}{2}\partial_i\sigma\right)^2=\partial^2-(D-2)(\partial_i\sigma)\partial^i-\frac{D-2}{2}(\partial^2\sigma)+\frac{(D-2)^2}{4}(\partial_i\sigma)(\partial^i\sigma) \label{partial term}
\end{equation}
Then, substituting (\ref{partial term}) for (\ref{Laplacian}), we arrive at the expression
\begin{eqnarray}
 \Delta &=& e^{2\sigma}\left[\left(\partial_i - \frac{D-2}{2}\partial_i\sigma \right)^2 + \frac{D-2}{2}(\partial^2\sigma)-\frac{(D-2)^2}{4}(\partial_i\sigma)(\partial^i\sigma) \right] \nonumber \\
 &=& e^{2\sigma}\left(\partial_i-\frac{D-2}{2}\partial_i\sigma \right)^2+\frac{D-2}{4(D-1)}R .
\end{eqnarray}
In particular, for $(x^i)=(x^\mu,y)$ and $\sigma=\sigma(y)$ with $D=5$, we have
\begin{equation}
 R = - e^{2\sigma}\left[ 8\ddot{\sigma}-12(\dot{\sigma})^2 \right]
\end{equation}
and
\begin{equation}
 \Delta = e^{2\sigma}\left[\partial_\mu\partial^\mu-\left(\partial_y-\frac{3}{2}\dot{\sigma} \right)^2 \right] + \frac{3}{16}R ,
\end{equation}
where the dot denotes differentiation with respect to $y$.

\section{First-Order-Loop Correction to the Inverse Propagator $(iG)^{-1}$}

The first-order-loop correction to the inverse propagator $(iG)^{-1}$, Eq.\,(\ref{Green function}), can be derived simply from the equation

\begin{eqnarray}
 K_1G_{12} &=& \int{\cal D}\tilde{\varphi}(K_1\tilde{\varphi}_1)\tilde{\varphi}_2=\int{\cal D}\tilde{\varphi}\tilde{\varphi}_2\left(-\frac{1}{i}\frac{\delta}{\delta\tilde{\varphi}_1}e^{iS_0}\right)e^{iS_I} \nonumber \\
 &=& \frac{1}{i}\delta_{12}-\frac{\lambda}{3!}\int{\cal D}\tilde{\varphi}\tilde{\varphi}_2(\rho_1\tilde{\varphi}_1^3)e^{iS_q} \nonumber \\
 &=& \frac{1}{i}\delta_{12}-\frac{\lambda}{3!}\rho_1\int{\cal D}\tilde{\varphi}\tilde{\varphi}_1^3\left(-K_2^{-1}\frac{1}{\delta\tilde{\varphi}_2}e^{iS_0} \right)e^{iS_I} \nonumber \\
 &=& \frac{1}{i}\delta_{12}-\frac{1}{i}\frac{\lambda}{2!}\rho_1(K^{-1})_{12} G^{(0)}_{11}+O(\lambda^2), \label{first order}
\end{eqnarray}
where $S_0$ and $S_I$ are the $\lambda$-independent part and $\lambda$-dependent part of $S_q$, respectively. Furthermore, we have written $G^{(0)}_{11}=-i(K^{-1})_{11}=\int{\cal D}\tilde{\varphi}\tilde{\varphi}_1^2e^{iS_0}$. Multiplying from the left of (\ref{first order}) by $iK_1^{-1}$, we can derive

\begin{equation}  iG_{12} = \left(K^{-1}+\frac{\lambda}{2!}K^{-1}\rho\bar{G}^{(0)}K^{-1} \right)_{12}+O(\lambda),
 \label{first order 2}
\end{equation}
where $\bar{G}^{(0)}=-i\bar{K}^{-1}$ is the diagonal matrix defined by $(\bar{G}^{(0)})_{ij}=G^{(0)}_{ii}\delta_{ij}$. Then, inverting the resultant equation (\ref{first order 2}), Eq.\,(\ref{Green function}) is immediately obtained.

\end{document}